\documentclass[twocolumn,superscriptaddress,showpacs,nofootinbib,preprintnumbers,amssymb, nobibnotes, aps, prd]{revtex4-2}
\usepackage[utf8]{inputenc}
\usepackage{graphicx}
\usepackage{latexsym,amsmath,amssymb,amsthm,lmodern,float,url,bbm,xspace}
\usepackage{natbib}
\usepackage{color}
\usepackage{microtype}
\usepackage{import}
\usepackage{bbold}
\usepackage[plain]{fancyref}
\usepackage{varioref}
\usepackage{slashed}
\usepackage{multirow}
\usepackage{tikz}
\usepackage{scrextend}
\usepackage{braket}
\usetikzlibrary{shapes}
\usetikzlibrary{positioning}
\usepackage[normalem]{ulem}
\usepackage{comment}
\usepackage{adjustbox}
\usepackage{tikz-feynman}
\usepackage{subcaption}
\usepackage{amsmath}
\usepackage{ulem}

\newcommand{\be}{\begin{equation}}
\newcommand{\ee}{\end{equation}}

\newcommand{\bea}{\begin{eqnarray}}
\newcommand{\eea}{\end{eqnarray}}

\usepackage[colorlinks=true,backref=false, linktocpage=true,
citecolor=blue,urlcolor=blue,linkcolor=blue,pdfpagemode=UseOutlines]{hyperref}
\hypersetup{%
  bookmarksnumbered=true,
  pdftitle = {},
  pdfsubject = {},
  pdfauthor = {},
  pdfkeywords = {}
}

\let\Re\undefined

\DeclareMathOperator{\Re}{\text{Re}}

\begin{document}
\title{Exploring new resonances with direct top flavor changing interactions} 
\author{Min Huang}
\email{huangmin@ihep.ac.cn}
\affiliation{Theoretical Physics Division, Institute of High Energy Physics, Chinese Academy of Sciences, Beijing 100049, China}
\affiliation{School of Physics, University of Chinese Academy of Sciences, Beijing 100049, China}
\author{Yandong Liu}
\email{ydliu@bnu.edu.cn}
\affiliation{Key Laboratory of Beam Technology of Ministry of Education, School of Physics and Astronomy, Beijing Normal University, Beijing, 100875, China}
\affiliation{Institute of Radiation Technology, Beijing Academy of Science and Technology, Beijing 100875, China}
\author{Hao Zhang}
\email{zhanghao@ihep.ac.cn}
\affiliation{Theoretical Physics Division, Institute of High Energy Physics, Chinese Academy of Sciences, Beijing 100049, China}
\affiliation{School of Physics, University of Chinese Academy of Sciences, Beijing 100049, China}
\affiliation{Center for High Energy Physics, Peking University, Beijing 100871, China}

\date{\today}

\begin{abstract}
In this work, we investigate three typical new physics resonances which couple to the standard model (SM) quarks via direct top-quark flavor-changing interactions. We identify the possible SMEFT operators electroweak scale and analyze their phenomenology. 
\end{abstract}

\maketitle

\section{Introduction}

Flavor-changing neutral current (FCNC) processes, which are highly suppressed in the particle physics standard model (SM) due to the well-known Glashow–Iliopoulos–Maiani (GIM) mechanism \cite{Glashow:1970gm}, are one of the most powerful indirect probes of new physics (NP) beyond the SM. For this reason, over the past decades, the FCNC interactions with ``light'' flavors have been stringently constrained by precisely observables in flavor physics. On the other hand, the FCNC effects in top-quark sector remains comparatively unexplored and attracts more and more attention recent years \cite{ATLAS:2022gzn,ATLAS:2022per,ATLAS:2023qzr,MorenoLlacer:2019dvy,ATLAS:2021amo,Zhang_2014,Ferreira:2008cj,Ferreira_2006,Coimbra:2008qp,Bach:2012fb,Andrea_2023,Cordero_Cid_2004,Degrande_2015,Khanpour_2017,Zhang:2013xya}. There are two main methods that are widely used in the analysis of the top-quark FCNC processes. One is starting from all possible effective operators in the SMEFT and treat them without any bias, which is well known as the model-independent method. The other is starting from specific NP models which contain heavy degree of freedoms that can induce FCNC processes in the top-quark sector.

The model-independent method, though is very generic and powerful for illustrating the results from experiments, keeps little information of the NP models behind the operators. To dig more information of the NP models from data, one needs to go a step forward and consider the operator correlations and show the relation between specific new particles and the specific operator correlation behavior. In this work, we consider a comprehensive set of heavy scalar and vector bosons that are consistent with SM gauge symmetry and couple to the top quark via renormalizable up-sector flavor changing effect interactions. Rather than treating SMEFT operators as arbitrary deformations, we derive them from explicit UV particle content, thereby enforcing correlations among operators that are otherwise invisible in purely model-independent analyses. This approach allows us to address questions that cannot be directly answered with generic SMEFT fits alone. 

The paper is organized as follows. In Sec. II, we construct and classify the viable heavy bosons and define their interactions with SM quarks. In Sec. III, we perform the matching to the SMEFT and discuss their LHC phenomenology. We conclude in Sec. IV with a discussion of the implications and outlook.

\section{The heavy resonances and the effective operators} 
\label{sec:UV}

\begin{table*}[!htbp]
\captionsetup{justification=raggedright, singlelinecheck=false}
  \caption{The quantum numbers and the interaction vertex of the heavy new particles\footnote{In this work, to make it clearly, we will use first several latin characters $a,b,c,\cdots$ denote , use $k,l,m,n\cdots$ denote the generation (flavor) of the SM fermions, use $A,B$ denote the color indices for the sextet and octet representations of the $SU(3)_c$ gauge group, use $\alpha, \beta,\cdots$ denote the color indices for the fundamental representation of the $SU(3)_c$ gauge group, use $i,j$ denote the indices for the fundamental representation of the $SU(2)_L$ gauge group, and use $I,J$ denote the indices for the adjoint representation of the $SU(2)_L$ gauge group.}.}
  \begin{tabular}{c|cccccc}
  \hline\hline
    ~~~~Name~~~~&~~~~ Spin ~~~~&~~~~ $\Delta F$ ~~~~&~~~~ $SU(3)_c$ ~~~~&~~~~ $SU(2)_L$ ~~~~&~~~~ $U(1)_Y$ ~~~~&~~~~
    Interaction\\
    \hline\hline
    $H_u$ & 0 & 0 & 1 & 2 & $- \frac{1}{2}$ & $y_{k l} \overline{Q_L}^{(k)}_{\alpha,i}u_{R}^{(l),\alpha}
    H_u^i + {\text{h.c.}}$\\
    \hline
    $H_d$ & 0 & 0 & 1 & 2 & $ \frac{1}{2}$ & $y_{k l} \overline{Q_L}^{(k)}_{\alpha,i}d_{R}^{(l),\alpha}
    H_d^i + {\text{h.c.}}$\\
    \hline
    $S_{u}$ & 0 & 0 & 8 & 2 & $- \frac{1}{2}$ & $y_{k l} \overline{Q_L}^{(k)}_{\alpha,i}(T^A)^{\alpha}_{~~\beta}u_{R}^{(l),\beta} S_{u,A}^i+ {\text{h.c.}}$\\
    \hline
    $S_{d}$ & 0 & 0 & 8 & 2 & $ \frac{1}{2}$ & $y_{k l} \overline{Q_L}^{(k)}_{\alpha,i}(T^A)^{\alpha}_{~~\beta}d_{R}^{(l),\beta} S_{d,A}^i + {\text{h.c.}}$\\
    \hline
    $(Z_R^\prime)_\mu$ & 1 & 0 & 1 & 1 & 0 & $y_{k l} \overline{u_R}_\alpha^{(k)}
    \gamma^\mu u_R^{(l),\alpha} (Z_R^\prime)_\mu$\\
    \hline
    $(G_R^\prime)_\mu$ & 1 & 0 & 8 & 1 & 0 & $y_{k l} \overline{u_R}_\alpha^{(k)}
    \gamma^\mu (T^A)^{\alpha}_{~~\beta} u_R^{(l),\beta} (G_R^\prime)_{A,\mu}$\\
    \hline
    $(Z_L^\prime)_\mu$ & 1 & 0 & 1 & 1 & 0 & $y_{k l} \overline{Q_L}_{\alpha,i}^{(k)}
    \gamma^\mu Q_L^{(l),\alpha,i} (Z_L^\prime)_\mu$\\
    \hline
    $(G_L^\prime)_\mu$ & 1 & 0 & 8 & 1 & 0 & $y_{k l} \overline{Q_L}_{\alpha,i}^{(k)}
    \gamma^\mu (T^A)^{\alpha}_{~~\beta} Q_L^{(l),\beta,i} (G_L^\prime)_{A,\mu}$\\
    \hline
    $(W^\prime_{L})_\mu$ & 1 & 0 & 1 & 3 & 0 & $y_{k l} \overline{Q_L}^{(k)}_{\alpha,i} 
    \gamma^{\mu} (\tau^I)^i_{~j} Q_L^{(l),\alpha,j} (W^\prime_{L})_{I,\mu}$\\
    \hline
    $(\Omega^\prime_{L})_\mu$ & 1 & 0 & 8 & 3 & 0 & $y_{k l} \overline{Q_L}^{(k)}_{\alpha,i} 
    \gamma^{\mu} (T^A)^{\alpha}_{~~\beta}(\tau^I)^i_{~j} Q_L^{(l),\alpha,j} (\Omega^\prime_{L})_{A,I,\mu}$\\
    \hline
    $\tilde S_{R}$ & 0 & 2 & 6 & 1 & $\frac{4}{3}$ & $y_{k l} K^{\alpha \beta}_A
    \overline{u_R}^{(k)}_\alpha (u^{(l)}_R)^C_\beta \tilde S_R^A + {\text{h.c.}}$\\
    \hline
    $\tilde S_{L}$ & 0 & 2 & 6 & 3 & $\frac{1}{3}$ & $y_{k l} K^{\alpha \beta}_A
    \overline{Q_L}^{(k)}_{\alpha,i} (\tau^I)^i_{~j} (Q^{(l)}_L)^C_{\beta,j} \tilde S_R^{A,I} + {\text{h.c.}}$\\
    \hline
    $V^\prime_\mu$ & 1 & 2 & 6 & 2 & $\frac{5}{6}$ & $y_{k l} K^{\alpha
    \beta}_A \overline{Q_L}^{(k)}_{\alpha,i} \gamma^{\mu} (u^{(l)}_R)^C_\beta V_\mu^{\prime A,i}
    + {\text{h.c.}}$\\
    \hline\hline
  \end{tabular}
  \label{tab:particles}
\end{table*}

In this work, we consider both spin-0 and spin-1 new particles in the NP models. For simplicity, we add some requirements on them and the renormalizable new interactions in our discussion as following:
\begin{itemize}
    \item[(1)] the new particle must has well-defined quantum number under the SM gauge group;
    \item[(2)] the new particle must couple to the top-quark with direct top-flavor-changing-interaction;
    \item[(3)] there is no flavor conserved interactions between the new particle and the SM quarks;
    \item[(4)] the exotic particle couples to the SM quarks via only one type of interaction. \footnote{This means that if a neutral vector boson couples to the right-handed up-type quark current, we will not assume it also couples to the right-handed down-type quark current directly.}
\end{itemize}

Imposing these requirements, we list the possible new particles and interactions in Table~\ref{tab:particles}, where $Q_L$ denotes the left-handed quark doublet, $u_R$ and $d_R$ are the right-handed singlets,  $T^a$ and $\tau^i$ are the generators of $SU(3)_c$ and $SU(2)$ in the fundamental representation, respectively, $K_{\alpha \beta}^a$ is the Clebsch-Gordan (CG) coefficient for the symmetric coupling of two color triplets to a sextet, and $\psi^c$ is the charge-conjugation of the fermion field $\psi$. One probably has noticed that the $Z_{R 1 \mu}^{{}'}(Z_{R 8 \mu}^{{}'})$ and $Z_{L 1 \mu}^{{}'}(Z_{L 8 \mu}^{{}'})$ have same quantum numbers, which means that the new fermion number conserved color singlet or octet vector bosons can couple to both the left-handed and the right-handed quarks. However, in this work we will consider one kind of interactions at one time for simplicity again. 

With these requirements, there remains some complexity of the flavor structure, which is important for inducing top-quark FCNC processes. To avoid additional contribution to the CP-violation, which is out of the scope of this work, we will assume $y_{ij}=y_{ji}\in {\mathbf{R}}$ for simplicity since now, and discuss the flavor structure for different exotic particles in detail. With this assumption, there are 3 independent parameters, $y_{12}$, $y_{23}$, and $y_{13}$ for each case. However, if all of these 3 parameters are non-zero, a simple triangle loop diagram would generate flavor diagonal interactions for the exotic particle. So we will open 2 of them at one time. Moreover, since any new interactions with bottom quark would get strong constraints from flavor physics observables, we will not consider the new resonances those couple to left-handed quark field. So we will only investigate the collider phenomenology in detail of the $Z_R^\prime$, $G_R^\prime$ and $\tilde S_R$. To get the correct combination and relative strength of the effective operators at the electroweak (EW) energy scale corresponding to the specific heavy particle, we first integrate out the heavy particle at its mass (cutoff) scale $\Lambda$ with the covariant derivative expansion (CDE) approach \cite{Cheyette:1985ue} to get the dim-6 operators in the Warsaw basis in SMEFT \cite{Grzadkowski:2010es} and their Wilson coefficients with the \texttt{MatchingTools} package~\cite{Criado:2017khh}. The relevant operator at the cutoff scale is
\begin{equation}
O_{uu,ijkl}=(\bar u_i\gamma_\mu u_j)(\bar u_k\gamma^\mu u_l).
\end{equation} 
The Wilson coefficients are 
\begin{eqnarray}
C_{uu,ijkl}(Z_R^\prime)&=&-y_{ij}y_{kl},\\
C_{uu,ijkl}(G_R^\prime)&=&-\frac{1}{2}y_{il}y_{kj}+\frac{1}{6}y_{ij}y_{kl},\\
C_{uu,ijkl}(\tilde S_R)&=&\frac{1}{4}y_{ik}y_{jl}.
\end{eqnarray}
After that we and evolve them down to the top-quark mass scale $m_t$ with the renormalization group equations (RGE),
\begin{equation}
    \frac{d C_i(\mu)}{d \log \mu} = \sum_j \frac{1}{16\pi^2} \gamma_{ij}\, C_j(\mu)\,,
    \label{eq:mixture}
\end{equation}
where $\gamma_{ij}$ is the anomalous-dimension matrix~\cite{Jenkins:2013wua,Jenkins:2013zja,Alonso:2013hga}. During the running, additional operators can be generated via mixing. Although these mixing-induced operators, comparing with the tree-level induced four-fermion operators, are suppressed by a loop factor, they could play some roles and should be checked. We focus in particular on the operators $O_{\varphi u}$, $O_{u\varphi}$, $O_{\varphi q}^{(1)}$, and $O_{\varphi q}^{(3)}$, which contribute directly to $t \to u (c) + Z/h$ processes~\cite{Aguilar_Saavedra_2009,Zhang_2011,BessidskaiaBylund:2016jvp}. The explicit forms of these operators are
\begin{eqnarray}
  O_{\varphi u} &=& (\varphi^\dagger i \overleftrightarrow{D}_\mu \varphi)(\bar{u}_i \gamma^\mu u_j)\,, \nonumber \\
  O_{\varphi q}^{(1)} &=& (\varphi^\dagger i \overleftrightarrow{D}_\mu \varphi)(\bar{q}_i \gamma^\mu q_j)\,, \nonumber \\
  O_{\varphi q}^{(3)} &=& (\varphi^\dagger i \overleftrightarrow{D}_\mu^i \varphi)(\bar{q}_i \tau^i \gamma^\mu q_j)\,,  \label{eq:rgeff}\\
  O_{u\varphi} &=& (\varphi^\dagger \varphi)(\bar{q}_i u_j \tilde{\varphi}),\nonumber
\end{eqnarray}
where either $i$ or $j$ gives a top-quark field operator. The partial width of top quark decay processes induced by these operators are
\bea
\Gamma(t\to qZ)&=&\frac{m_t^3}{32\sqrt2\pi G_F\Lambda^4}\left(1-\frac{m_Z^2}{m_t^2}\right)^2\left(1+2\frac{m_Z^2}{m_t^2}\right)\nonumber\\
&&\times\left(|C_{\phi q}^{(3)}-C_{\phi q}^{(1)}|^2+|C_{\phi u}|^2\right)\nonumber\\
&=&\frac{0.0019}{\Lambda^4/(1{\text{TeV}})^4}\Gamma_t\left(|C_{\phi q}^{(3)}-C_{\phi q}^{(1)}|^2+|C_{\phi u}|^2\right),\nonumber\\
\\
\Gamma(t\to qh)&=&\frac{m_t}{128\pi G_F^2\Lambda^4}\left(1-\frac{m_h^2}{m_t^2}\right)^2|C_{u\phi}|^2\nonumber\\
&=&\frac{0.00054}{\Lambda^4/(1{\text{TeV}})^4}\Gamma_t|C_{u\phi}|^2.
\eea
Setting the coupling coefficients $y_{ij}$ and cutoff scales, we employ the \texttt{Wilson} package~\cite{Aebischer_2018_2} to compute the these running Wilson coefficients at the top-quark mass scale. For simplicity, we will not run the tree-level Wilson coefficients $C_{uu,ijkl}$'s in this work.

\section{The Collider Phenomenology} 
\label{sec:pheno} 

In this section we investigate the phenomenology associated with the effective operators derived in Sec.~\ref{sec:UV} and analyze the different behavior of the three different new resonances. For each new particles, we assign the theoretical inputs at the cutoff scale, where only four fermion operators are generated. At the electroweak (EW) scale, two kinds of operators are phenomenologically important: the four-fermion interaction $O_{uu}$, and the purely RG-induced operators listed in Eq. (\ref{eq:rgeff}). The four-fermion operators probably contribute to the single top-quark production processes, the branching ratio of top quark decaying to 3 jets process, the pair production of both same-sign and opposite-sign top-quarks at colliders significantly \cite{Aguilar-Saavedra:2010uur}, while the RG-induced operators induced both single top-quark production signal and the anomalous decay signal of the top quark, which could be searched at a ``top-factory'' such as the LHC. On the other hand, the RG-induced operators would also predict some anomalous production channel at Higgs factory \cite{Shi:2019epw}. We check the main the phenomenological constraints of these two kinds of operators at the LHC. All of the signal cross sections are generated with \texttt{MadGraph5\_aMC@NLO} \cite{Alwall:2014hca} to leading order (LO) at parton level with the NNPDF2.3 LO parton distribution function (PDF) \cite{Ball:2013hta}. 

We separate the combinations of the FCNC coupling constants into 3 different patterns, $y_{23}=0$, $y_{31}=0$, and $y_{12}=0$. Before going into the details of the analysis of the phenomenology, we first list the characters of different patterns.

\subsection{The three different patterns}
We first list and analyze the three different patterns.
\subsubsection{$y_{23}=0$}
The effective operators can be separated into three groups:
\begin{description}
\item[ Group-1 ] The operators in this group contain one (anti)top-quark field operator. So they contribute to the single top-quark production and top-quark rare decay ($t\to jjj$) process. They are (we denote $O_{\bar ij\bar kl}$ for $O_{uu,ijkl}$ since now for simplicity. )
\begin{eqnarray}
\mathcal{H}_{Z^\prime_R}&=&-\frac{y_{12}y_{13}}{\Lambda^2}(O_{\bar uc\bar ut}+O_{\bar cu\bar ut}+O_{\bar uc\bar tu}+O_{\bar cu\bar tu}),\nonumber\\
\mathcal{H}_{G^\prime_R}&=&-\frac{y_{12}y_{13}}{3\Lambda^2}(O_{\bar uc\bar ut}+O_{\bar cu\bar tu})\nonumber\\
&&~~~~+\frac{y_{12}y_{13}}{6\Lambda^2}(O_{\bar cu\bar ut}+O_{\bar uc\bar tu})\nonumber\\
&&~~~~~~~~-\frac{y_{12}y_{13}}{2\Lambda^2}(O_{\bar uu\bar tc}+O_{\bar uu\bar ct}),\nonumber\\
\mathcal{H}_{\tilde S_R}&=&\frac{y_{12}y_{13}}{4\Lambda^2}(O_{\bar uu\bar ct}+O_{\bar cu\bar ut}+O_{\bar uu\bar tc}+O_{\bar uc\bar tu}).\nonumber\\
\end{eqnarray}
Since there is no interference between the SM process and the effective operators, the leading order contribution to the production cross section of single top quark can be factorized as
\begin{eqnarray}
\delta\sigma(pp\to tj)&=&\delta\sigma_t^{(1)}\frac{S^2}{\Lambda^4}y_{12}^2y_{13}^2,\\
\delta\sigma(pp\to \bar tj)&=&\delta\sigma_{\bar t}^{(1)}\frac{S^2}{\Lambda^4}y_{12}^2y_{13}^2,
\end{eqnarray}
where $\sqrt{S}=1$TeV for simplicity.
\item[ Group-2 ] The operators in this group contain two top-quark field operators. So they contribute to the top-quark pair production and the same-sign top-quark pair production processes. They are
\begin{eqnarray}
\mathcal{H}_{Z^\prime_R}&=&-\frac{y_{13}^2}{\Lambda^2}(O_{\bar ut\bar ut}+O_{\bar ut\bar tu}+O_{\bar tu\bar tu}),\nonumber\\
\mathcal{H}_{G^\prime_R}&=&-\frac{y_{13}^2}{3\Lambda^2}(O_{\bar ut\bar ut}+O_{\bar tu\bar tu})\nonumber\\
&&~~~~+\frac{y_{13}^2}{6\Lambda^2}O_{\bar ut\bar tu}-\frac{y_{13}^2}{2\Lambda^2}O_{\bar uu\bar tt},\nonumber\\
\mathcal{H}_{\tilde S_R}&=&\frac{y_{13}^2}{4\Lambda^2}(O_{\bar uu\bar tt}+O_{\bar tu\bar ut}).
\end{eqnarray}
Note that the color sextet scalar will not lead to the exotic same-sign top-quark pair production (which means same-sign up-quark in the operator) at the LHC since it is a complex field which can be assigned a ``up-number'' $U(1)$ quantum number to restore the conservation of the up quark number in the Lagrangian in the UV theory.

Since the leading order contribution to the production cross section of top quark production is again the interference terms, one has
\begin{equation}
\delta\sigma(pp\to t\bar t)=\delta\sigma_{t\bar t}^{(1)}\frac{S}{\Lambda^2}y_{13}^2.
\end{equation}
Since there is no same-sign top quark pair process in the SM, the leading order contribution to the process is again $\mathcal{O}(\Lambda^{-4})$,
\begin{eqnarray}
\delta\sigma(pp\to tt)&=&\delta\sigma_{tt}^{(1)}\frac{S^2}{\Lambda^4}y_{13}^4,\\
\delta\sigma(pp\to \bar t\bar t)&=&\delta\sigma_{\bar t\bar t}^{(1)}\frac{S^2}{\Lambda^4}y_{13}^4.
\end{eqnarray}
\item[ Group-3 ]  The operators in this group do not contain any top-quark field operator. But they probably contribute to the $D^0-\bar D^0$ mixing. They are
\begin{eqnarray}
\mathcal{H}_{Z^\prime_R}&=&-\frac{y_{12}^2}{\Lambda^2}(O_{\bar uc\bar uc}+O_{\bar uc\bar cu}+O_{\bar cu\bar cu})\nonumber\\
&\to&-\frac{y_{12}^2}{\Lambda^2}(O_{\bar uc\bar uc}+O_{\bar cu\bar cu}),\nonumber\\
\mathcal{H}_{G^\prime_R}&=&-\frac{y_{12}^2}{3\Lambda^2}(O_{\bar uc\bar uc}+O_{\bar cu\bar cu})\nonumber\\
&&~~~~+\frac{y_{12}^2}{6\Lambda^2}O_{\bar uc\bar cu}-\frac{y_{12}^2}{2\Lambda^2}O_{\bar uu\bar cc}\nonumber\\
&\to&-\frac{y_{12}^2}{3\Lambda^2}(O_{\bar uc\bar uc}+O_{\bar cu\bar cu}),\nonumber\\
\mathcal{H}_{\tilde S_R}&=&\frac{y_{12}^2}{4\Lambda^2}(O_{\bar uu\bar cc}+O_{\bar cu\bar uc})\to 0.
\end{eqnarray}
Again, the color sextet scalar does not give any contribution due to the up-number conservation.
\end{description}

\subsubsection{$y_{13}=0$}
The phenomenology of this case is nearly the same to the first case except for the difference between the parton distribution of the up-quark and the charm-quark. The interaction Hamiltonian and the corrections in the three groups are
\begin{description}
\item[ Group-1 ] The operators are
\begin{eqnarray}
\mathcal{H}_{Z^\prime_R}&=&-\frac{y_{12}y_{23}}{\Lambda^2}(O_{\bar uc\bar ct}+O_{\bar cu\bar ct}+O_{\bar uc\bar tc}+O_{\bar cu\bar tc}),\nonumber\\
\mathcal{H}_{G^\prime_R}&=&-\frac{y_{12}y_{23}}{3\Lambda^2}(O_{\bar uc\bar ct}+O_{\bar cu\bar tc})\nonumber\\
&&~~~~+\frac{y_{12}y_{23}}{6\Lambda^2}(O_{\bar cu\bar ct}+O_{\bar uc\bar tc})\nonumber\\
&&~~~~~~~~-\frac{y_{12}y_{23}}{2\Lambda^2}(O_{\bar cc\bar tu}+O_{\bar cc\bar ut}),\nonumber\\
\mathcal{H}_{\tilde S_R}&=&\frac{y_{12}y_{23}}{4\Lambda^2}(O_{\bar uc\bar ct}+O_{\bar cc\bar ut}+O_{\bar cc\bar tu}+O_{\bar cu\bar tc}).\nonumber\\
\end{eqnarray}
The coefficients for single top quark production are
\begin{eqnarray}
\delta\sigma(pp\to tj)&=&\delta\sigma_t^{(2)}\frac{S^2}{\Lambda^4}y_{12}^2y_{23}^2,\\
\delta\sigma(pp\to \bar tj)&=&\delta\sigma_{\bar t}^{(2)}\frac{S^2}{\Lambda^4}y_{12}^2y_{23}^2.
\end{eqnarray}
\item[ Group-2 ] The operators are
\begin{eqnarray}
\mathcal{H}_{Z^\prime_R}&=&-\frac{y_{23}^2}{\Lambda^2}(O_{\bar ct\bar ct}+O_{\bar ct\bar tc}+O_{\bar tc\bar tc}),\nonumber\\
\mathcal{H}_{G^\prime_R}&=&-\frac{y_{23}^2}{3\Lambda^2}(O_{\bar ct\bar ct}+O_{\bar tc\bar tc})\nonumber\\
&&~~~~+\frac{y_{23}^2}{6\Lambda^2}O_{\bar ct\bar tc}-\frac{y_{23}^2}{2\Lambda^2}O_{\bar cc\bar tt},\nonumber\\
\mathcal{H}_{\tilde S_R}&=&\frac{y_{23}^2}{4\Lambda^2}(O_{\bar cc\bar tt}+O_{\bar tc\bar ct}).
\end{eqnarray}
Again the color sextet scalar will not lead to the exotic same-sign top-quark pair production dut to the charm-number conservation.

The coefficients for top quark pair and same-sign top quark pair productions are
\begin{eqnarray}
\delta\sigma(pp\to t\bar t)&=&\delta\sigma_{t\bar t}^{(2)}\frac{S}{\Lambda^2}y_{23}^2,\\
\delta\sigma(pp\to tt)&=&\delta\sigma_{tt}^{(2)}\frac{S^2}{\Lambda^4}y_{23}^4,\\
\delta\sigma(pp\to \bar t\bar t)&=&\delta\sigma_{\bar t\bar t}^{(2)}\frac{S^2}{\Lambda^4}y_{23}^4.
\end{eqnarray}
\item[ Group-3 ]  The operators in this group are exactly the same to the first case.
\end{description}

\subsubsection{$y_{12}=0$}
In this case, there is no contribution to the single top quark production processes. The relevant interaction Hamiltonians for the top quark pair production process are
\begin{eqnarray}
\mathcal{H}_{Z^\prime_R}&=&-\frac{y_{13}y_{23}}{\Lambda^2}(O_{\bar ut\bar tc}+O_{\bar tu\bar ct})-\frac{y_{13}^2}{\Lambda^2}O_{\bar t u\bar ut}\nonumber\\
&&~~-\frac{y_{23}^2}{\Lambda^2}O_{\bar t c\bar ct},\nonumber\\
\mathcal{H}_{G^\prime_R}&=&\frac{y_{13}y_{23}}{6\Lambda^2}(O_{\bar ut\bar tc}+O_{\bar tu\bar ct})\nonumber\\
&&~~-\frac{y_{13}y_{23}}{2\Lambda^2}(O_{\bar uc\bar tt}+O_{\bar cu\bar tt})\nonumber\\
&&~~~~+\frac{y_{13}^2}{6\Lambda^2}O_{\bar ut\bar tu}-\frac{y_{13}^2}{2\Lambda^2}O_{\bar uu\bar tt}\nonumber\\
&&~~~~~~+\frac{y_{23}^2}{6\Lambda^2}O_{\bar ct\bar tc}-\frac{y_{23}^2}{2\Lambda^2}O_{\bar cc\bar tt},\nonumber\\
\mathcal{H}_{\tilde S_R}&=&\frac{y_{13}y_{23}}{4\Lambda^2}(O_{\bar uc\bar tt}+O_{\bar ut\bar tc}+O_{\bar cu\bar tt}+O_{\bar tu\bar ct})\nonumber\\
&&~~+\frac{y_{13}^2}{4\Lambda^2}O_{\bar uu\bar tt}+\frac{y_{23}^2}{4\Lambda^2}O_{\bar cc\bar tt},
\end{eqnarray}
while the relevant interaction Hamiltonians for the same-sign top quark pair production process are
\begin{eqnarray}
\mathcal{H}_{Z^\prime_R}&=&-\frac{y_{13}y_{23}}{\Lambda^2}(O_{\bar ut\bar ct}+O_{\bar tu\bar tc})-\frac{y_{13}^2}{\Lambda^2}(O_{\bar ut\bar ut}+O_{\bar tu\bar tu})\nonumber\\
&&~~-\frac{y_{23}^2}{\Lambda^2}(O_{\bar ct\bar ct}+O_{\bar tc\bar tc})\nonumber\\
\mathcal{H}_{G^\prime_R}&=&-\frac{y_{13}y_{23}}{3\Lambda^2}(O_{\bar ut\bar ct}+O_{\bar tu\bar tc})-\frac{y_{13}^2}{3\Lambda^2}(O_{\bar ut\bar ut}+O_{\bar tu\bar tu})\nonumber\\
&&~~-\frac{y_{23}^2}{3\Lambda^2}(O_{\bar ct\bar ct}+O_{\bar tc\bar tc}),\nonumber\\
\mathcal{H}_{\tilde S_R}&=&0.
\end{eqnarray}
The color sextet in this case does not contribute to the same-sign top quark pair production process due to the ``top-number'' conservation in the UV model. Since the operators which are proportional to $y_{12}y_{13}$ do not interfere with the SM, to the leading order, we have 
\begin{eqnarray}
\delta\sigma(pp\to t\bar t)&=&\delta\sigma_{t\bar t}^{(1)}\frac{S}{\Lambda^2}y_{13}^2+\delta\sigma_{t\bar t}^{(2)}\frac{S}{\Lambda^2}y_{23}^2,\\
\delta\sigma(pp\to tt)&=&\delta\sigma_{tt}^{(1)}\frac{S^2}{\Lambda^4}y_{13}^4+\delta\sigma_{tt}^{(3)}\frac{S^2}{\Lambda^4}y_{13}^2y_{23}^2\nonumber\\
&&~~+\delta\sigma_{tt}^{(2)}\frac{S^2}{\Lambda^4}y_{23}^4,\\
\delta\sigma(pp\to \bar t\bar t)&=&\delta\sigma_{\bar t\bar t}^{(1)}\frac{S^2}{\Lambda^4}y_{13}^4+\delta\sigma_{\bar t\bar t}^{(3)}\frac{S^2}{\Lambda^4}y_{13}^2y_{23}^2\nonumber\\
&&~~+\delta\sigma_{\bar t\bar t}^{(2)}\frac{S^2}{\Lambda^4}y_{23}^4.
\end{eqnarray}

The contribution to the $D^0-\bar D^0$ mixing in this case is suppressed by another factor of $(16\pi^2\Lambda^2)^{-1}$. The coefficients appear in the cross section of the top quark production processes at 13TeV LHC are listed in TABLE. \ref{tab:top}.

\begin{table*}[!htbp]
\captionsetup{justification=raggedright, singlelinecheck=false}
  \caption{The coefficients in the cross section of the top quark production processes at 13TeV LHC. The number shown in this table is in the unit of pb.}
  \begin{tabular}{c|c|c|c|c|c|c|c|c|c|c|c|c}
  \hline\hline
  & $\delta\sigma_t^{(1)}$ & $\delta\sigma_{\bar t}^{(1)}$ & $\delta\sigma_t^{(2)}$ & $\delta\sigma_{\bar t}^{(2)}$ & $\delta\sigma_{t\bar t}^{(1)}$ & $\delta\sigma_{tt}^{(1)}$ & $\delta\sigma_{\bar t\bar t}^{(1)}$ & $\delta\sigma_{t\bar t}^{(2)}$ & $\delta\sigma_{tt}^{(2)}$ & $\delta\sigma_{\bar t\bar t}^{(2)}$  & $\delta\sigma_{tt}^{(3)}$ & $\delta\sigma_{\bar t\bar t}^{(3)}$  \\
  \hline\hline
~~~~$Z^\prime_R$~~~~&~~365.4~~~&~~56.17~~~&~~30.57~~~&~~33.21~~~&~~-15.27~~~&~~517.3~~~&~~5.765~~~&~~-0.6751~~~&~~ 1.683~~~&~~1.683~~~&~~17.73~~~&~~1.558~~~\\
  \hline
~~$G^\prime_R$~~ &42.73 &7.926 &0.8497 &5.012 & 1.385& 57.58 &0.6412 & 0.0618& 0.1868&0.1868& 1.969& 0.1733\\
\hline
~~$\tilde S_R$~~ & 6.08&5.281 &1.657 & 0.5519& 4.394& $0$ & $0$ & 0.1941& $0$ & $0$ & 0& 0 \\
    \hline\hline
  \end{tabular}
  \label{tab:top}
\end{table*}

\subsection{The constraint from neutral $D$-meson system} 
\label{sec:d0d0bar}
Before going to the collider phenomenology, let us first check the constraint from the $D^0-\bar D^0$ mixing effect. The $|\Delta C=2|$ Hamiltonian is 
\begin{equation}
\mathcal H_{\Delta C=2}=\frac{C_{12}}{\Lambda^2}O_{\bar uc\bar uc},
\end{equation}
where $C_{12}$ is the Wilson coefficient at the electroweak scale. So the new physics contribution to $\Delta m_D\equiv|m_{D^0}-m_{\bar D^0}|$ is
\bea
\Delta m_D&=&\left|\Re\left(\frac{2\langle \bar D^0|\mathcal H|D^0\rangle}{2m_D}\right)\right|\nonumber\\
&=&\frac{1}{m_D}\frac{C_{12}}{\Lambda^2}|\langle O_{\bar uc\bar uc}\rangle|,
\eea
where the hadron matrix element is \cite{Golowich:2007ka}
\be
\langle O_{\bar uc\bar uc}\rangle=\frac{2}{3}f_D^2m_D^2B_D,
\ee
with $f_D=211.5\pm2.3\pm1.1\pm0.8$ MeV \cite{BESIII:2024kvt}, $m_D=1864$ MeV, and $B_D=0.82$. The measured result of $\Delta m_D$ is \cite{ParticleDataGroup:2024cfk}
\be
\Delta m_D=(6.2\pm0.72)\times10^{-12}{\text{MeV}},
\ee
which gives an roughly constraint to the Wilson coefficient and the cutoff scale:
\be
\frac{C_{12}}{\Lambda^2}\lesssim1.3\times 10^{-7}{\text{TeV}}^{-2}.
\ee
Although in a more carefully calculation, the RG running effect of the effective operators should be included, it will not qualitatively change the conclusion since the non-perturbative parameters used in this calculation contain large uncertainty and dominant the theoretical error. Thus, for the vector bosons $Z_R^\prime$ and $G_R^\prime$, $y_{12}/\Lambda^2$ is highly constrained by the $D^0-\bar D^0$ mixing effect. On the other hand, the color sextet does not get strong constraint from this observable at leading order.

However, the color sextet with $y_{12}=0$ can be hardly constrained by any other process. It is worth to use this observable to constrain the parameters. The contribution is from the loop induced contribution to the dimension-6 effective operator $O_{\bar uc\bar uc}$ which is $\mathcal{O}(\Lambda^{-4})$, and the limit is roughly
\be
\frac{y_{13}^2y_{23}^2m^2_t}{16\pi^2\Lambda^4}\lesssim1.3\times 10^{-7}{\text{TeV}}^{-2}.
\ee
Here we have ignore the extra logarithm factor $\log(\Lambda^2/m_t^2)$ since its correction is $\mathcal{O}(1)$ for a large range of $\Lambda$. This will not affect the result qualitatively, since the result is mostly suffered by the theoretical uncertainty from non-perturbative QCD.

\subsection{The LHC phenomenology} 
\label{sec:lhc_analysis}
For the single top-quark production channel, the signal events will contribute to the $t$-channel single top-quark production. The most precisely result is given in \cite{ATLAS:2024ojr}
\be
\sigma(tq)=137^{+8}_{-8}{\text{pb}},~~~~\sigma(\bar tq)=84^{+6}_{-5}{\text{pb}}.
\ee
One should keep in mind that here we only give a very roughly estimation due to several reasons:
\begin{itemize}
\item The different combination of the initial state partons will not only change the total cross section, but also change the rapidity of the $tq$ system. For example, there will be more signal events in the central region for the $uu$ initial state.
\item The top quark produced by the new physics model is right handed. So the charged lepton from its decay will be more energetic than the charged lepton from a left handed top quark. This fact means that the cut acceptance of the new physics single top quark event will be a little higher than the SM $t$-channel single top quark. However, this conclusion depends on the details of either the kinetic cuts or the weight parameters in the artificial neural networks used in the analysis.
\end{itemize}
Nevertheless, a fully detector level simulation is out of the scope of this work, for this channel and also the opposite-sign and same-sign top quark pair production channels. We leave these to future works. The SM predictions with NNLO QCD correction are \cite{Berger:2016oht}
\be
\sigma(tq)=134.3^{+1.3}_{-0.7}{\text{pb}},~~~~\sigma(\bar tq)=79.3^{+0.8}_{-0.5}{\text{pb}}.
\ee
At 95\% CL, we find that 
\be
\delta\sigma_t\leqslant18.6{\text{pb}},~~~~\delta\sigma_{\bar t}\leqslant16.6{\text{pb}}.
\ee

For the $t\bar t$ pair production channel, the total cross section from LHC is \cite{ATLAS:2023gsl}
\be
\sigma_{t\bar t}=829\pm15{\text{pb}},
\ee
while the SM prediction is \cite{ATLAS:2024kxj}
\be
\sigma_{t\bar t}=834^{+38}_{-43}{\text{pb}}.
\ee
At 95\% CL, we have
\be
-94{\text{pb}}\leqslant\delta\sigma_{t\bar t}\leqslant75{\text{pb}}.
\ee
Comparing with the numbers in TABLE \ref{tab:top}, one may find that the single top channel is more sensitive to the new physics discussed here than the $t\bar t$ channel.

The same-sign top quark pair production is another important channel for the new resonances discussed in this work. For this channel, the observed upper limit of $\sigma(tt)$ is 1.6fb at 95\% CL \cite{ATLAS:2024hac} \footnote{The constraint to $C_{\bar tu\bar tu}/\Lambda^2$ in that work can not be used directly here, since this is the result with the assumption of flavor universality. }. It is obviously that the strongest constraint to the $Z^\prime_R$ and $G^\prime_R$ would be from the $tt$ channel.

\begin{figure*}[htbp]
\begin{center}
\includegraphics[scale=0.25]{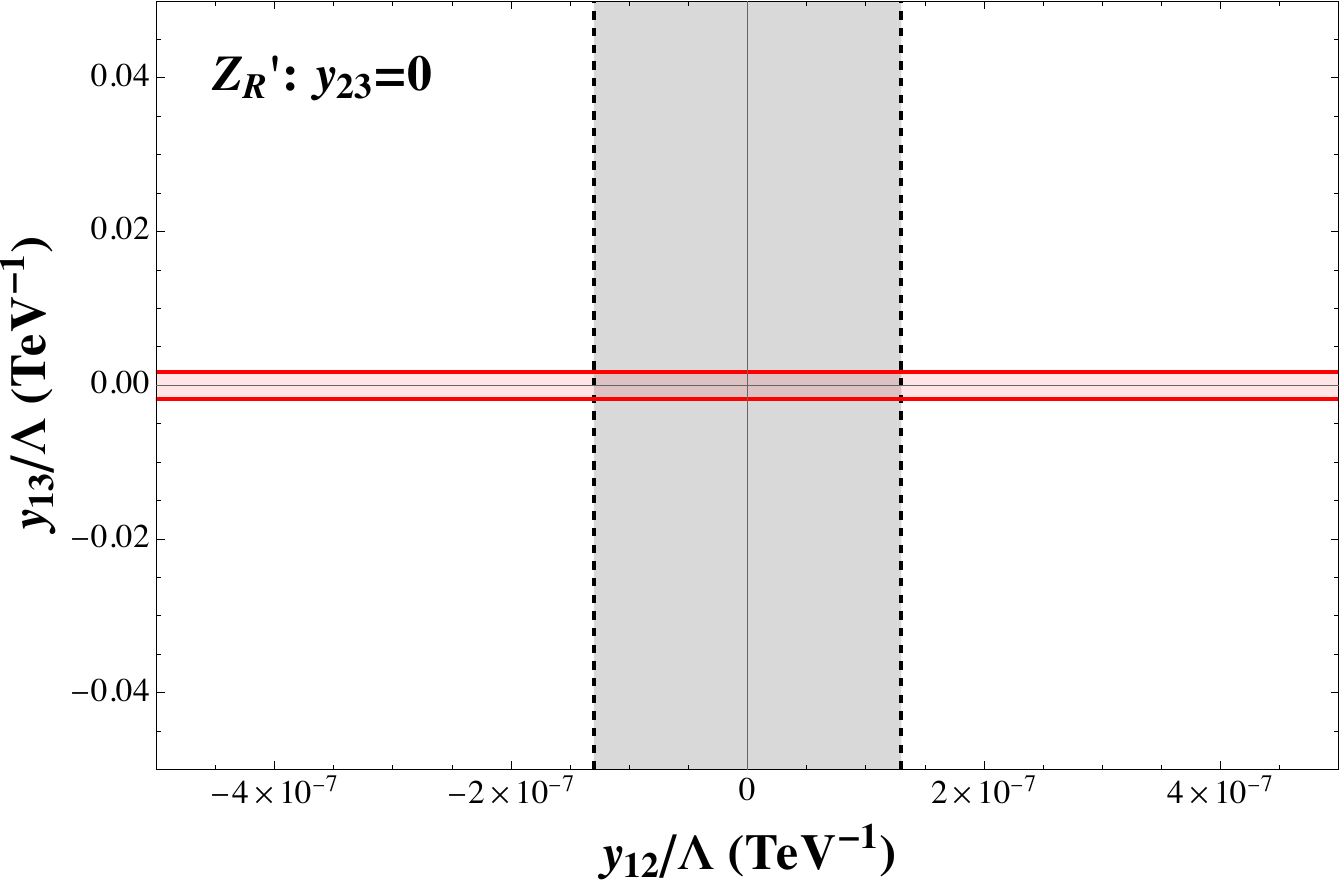}
\includegraphics[scale=0.25]{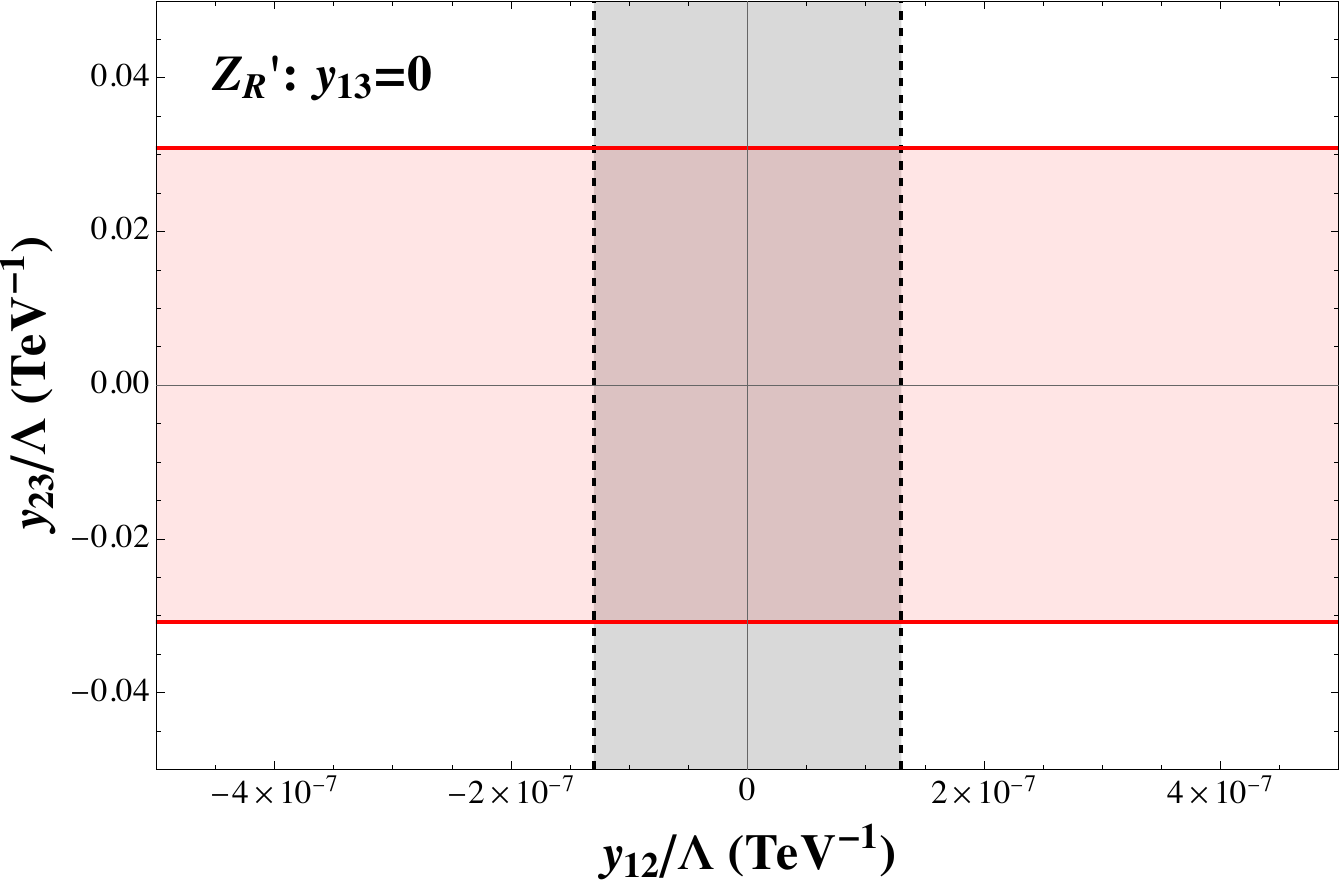}
\includegraphics[scale=0.25]{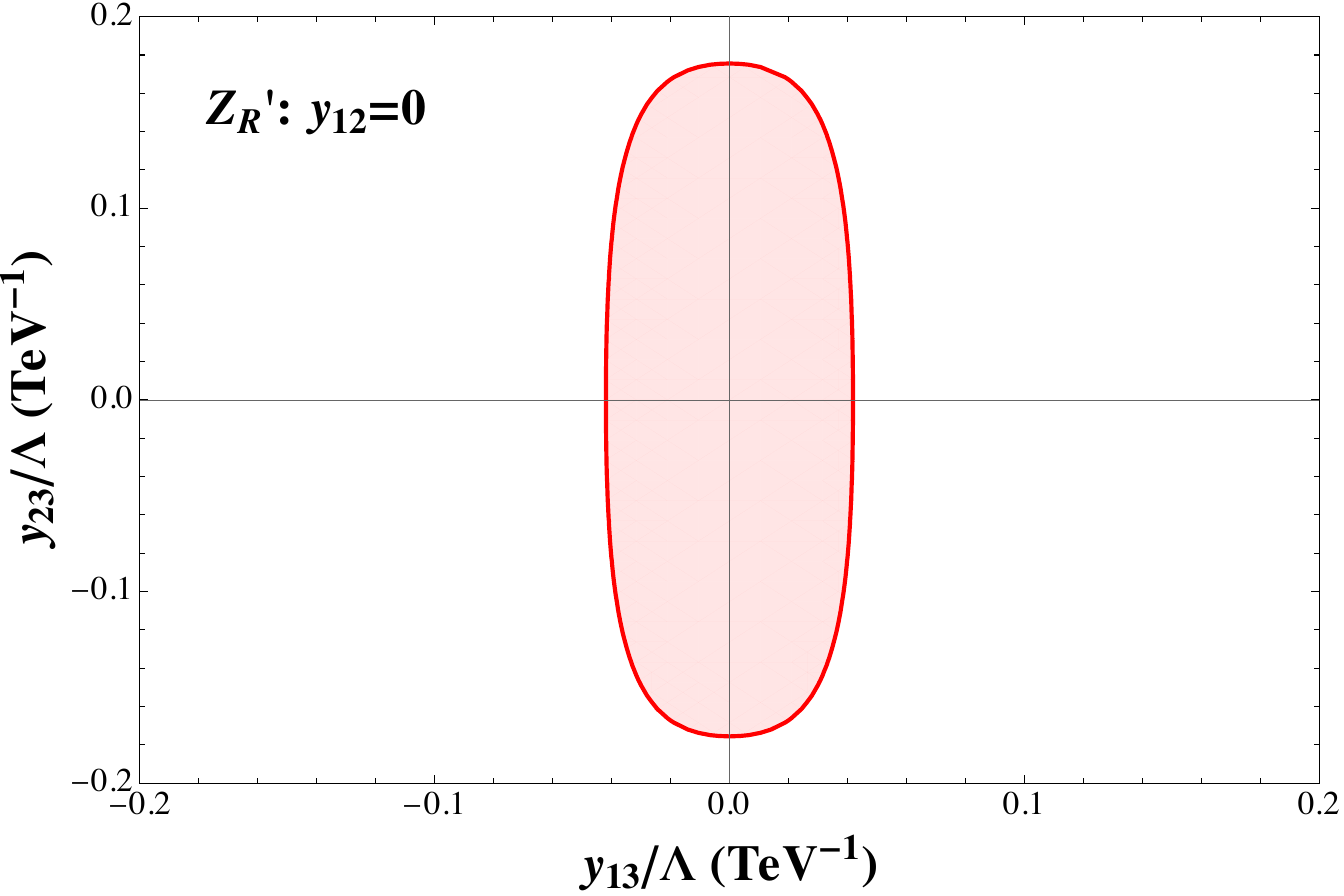}
\caption{The allowed regions in the parameter space for $Z^\prime_R$. The light red region is given by the same-sign top quark production cross section at 13TeV LHC. The gray region is given by the $D^0-\bar D^0$ mixing observable. The constraint from $D^0-\bar D^0$ mixing is not shown if it is always weak than the constraint from top physics. } 
\label{fig:zr}
\end{center}
\end{figure*}

\begin{figure*}[htbp]
\begin{center}
\includegraphics[scale=0.25]{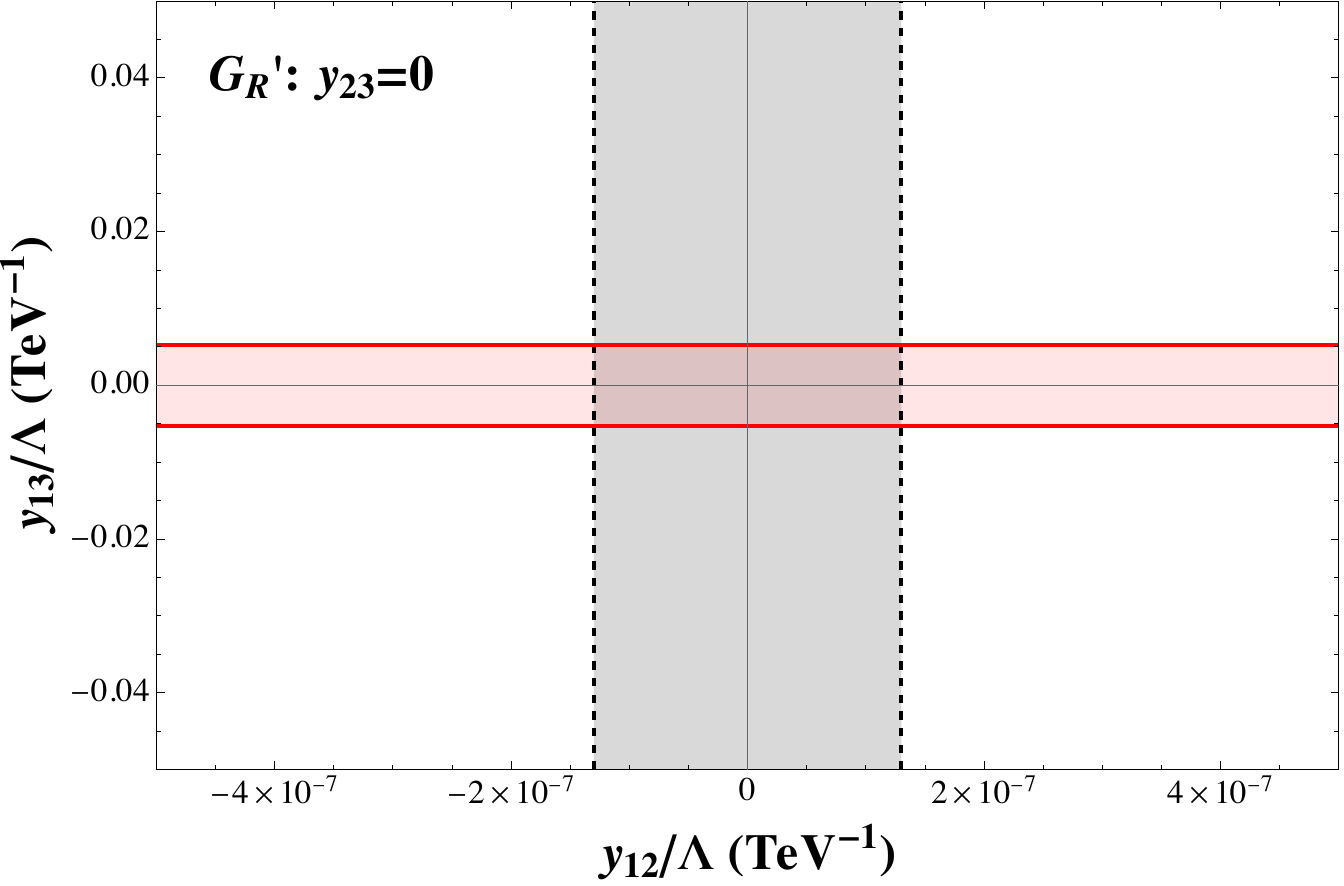}
\includegraphics[scale=0.25]{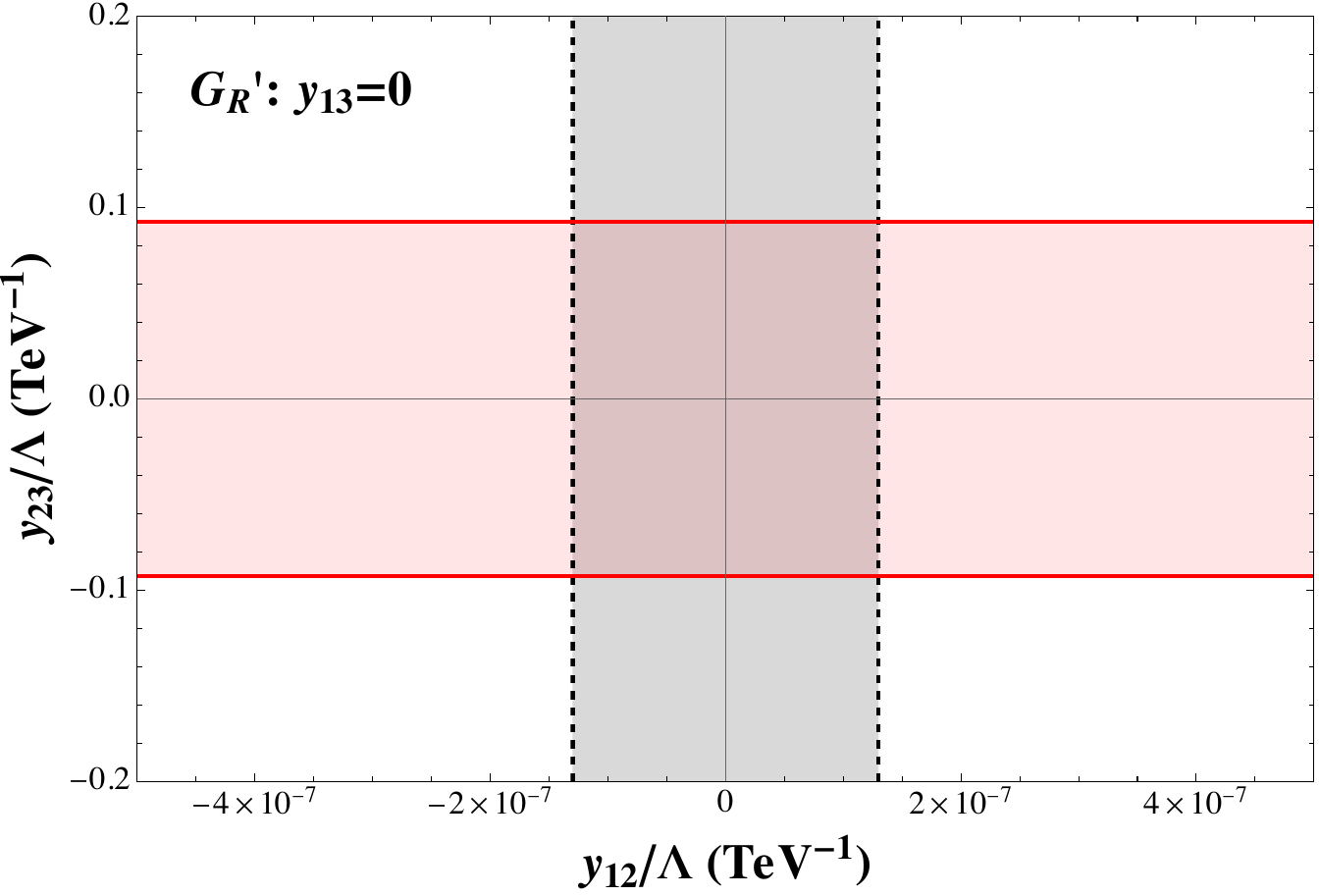}
\includegraphics[scale=0.25]{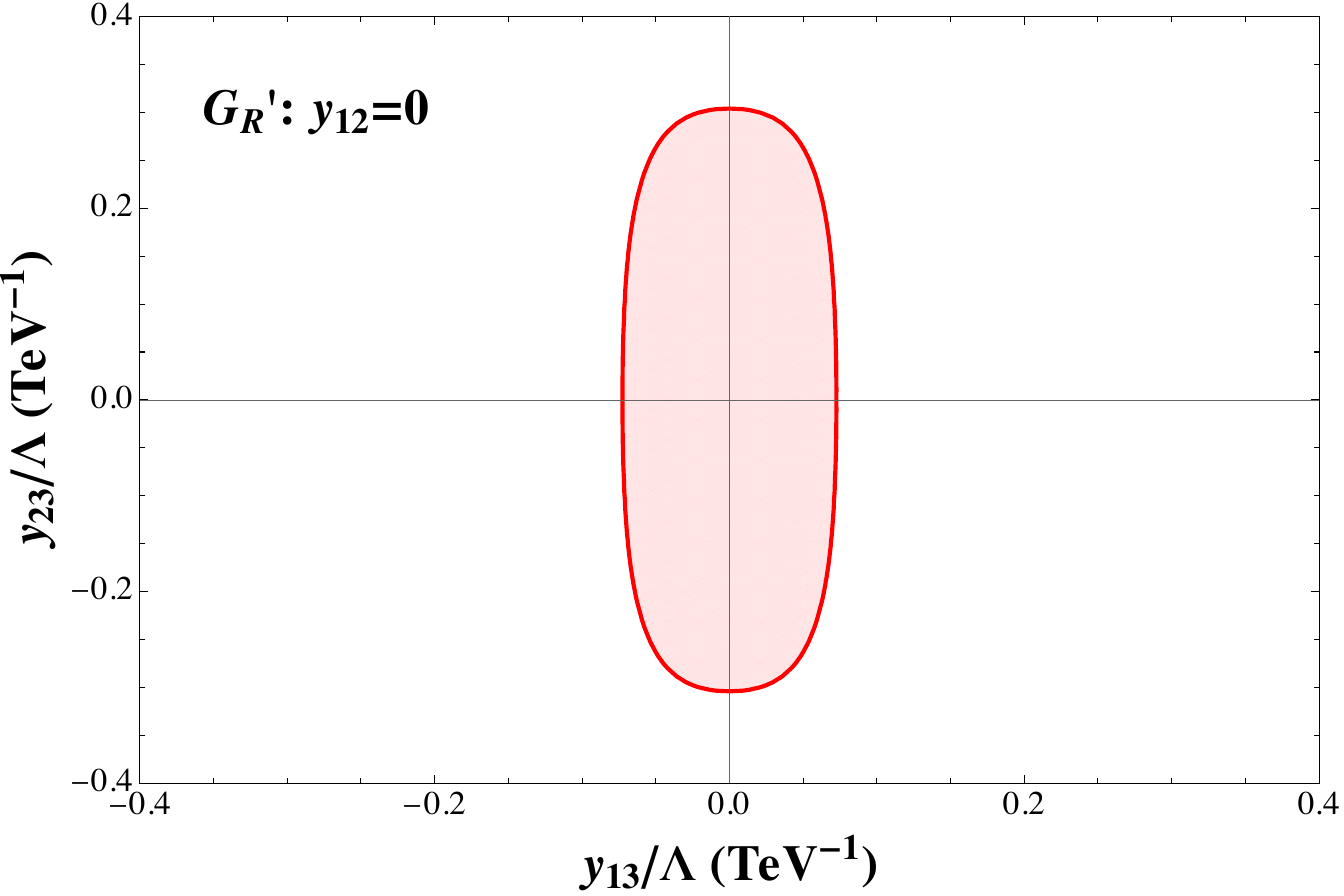}
\caption{The allowed regions in the parameter space for $G^\prime_R$. The light red region is given by the same-sign top quark production cross section at 13TeV LHC. The gray region is given by the $D^0-\bar D^0$ mixing observable. The constraint from $D^0-\bar D^0$ mixing is not shown if it is always weak than the constraint from top physics. } 
\label{fig:gr}
\end{center}
\end{figure*}

\begin{figure*}[htbp]
\begin{center}
\includegraphics[scale=0.25]{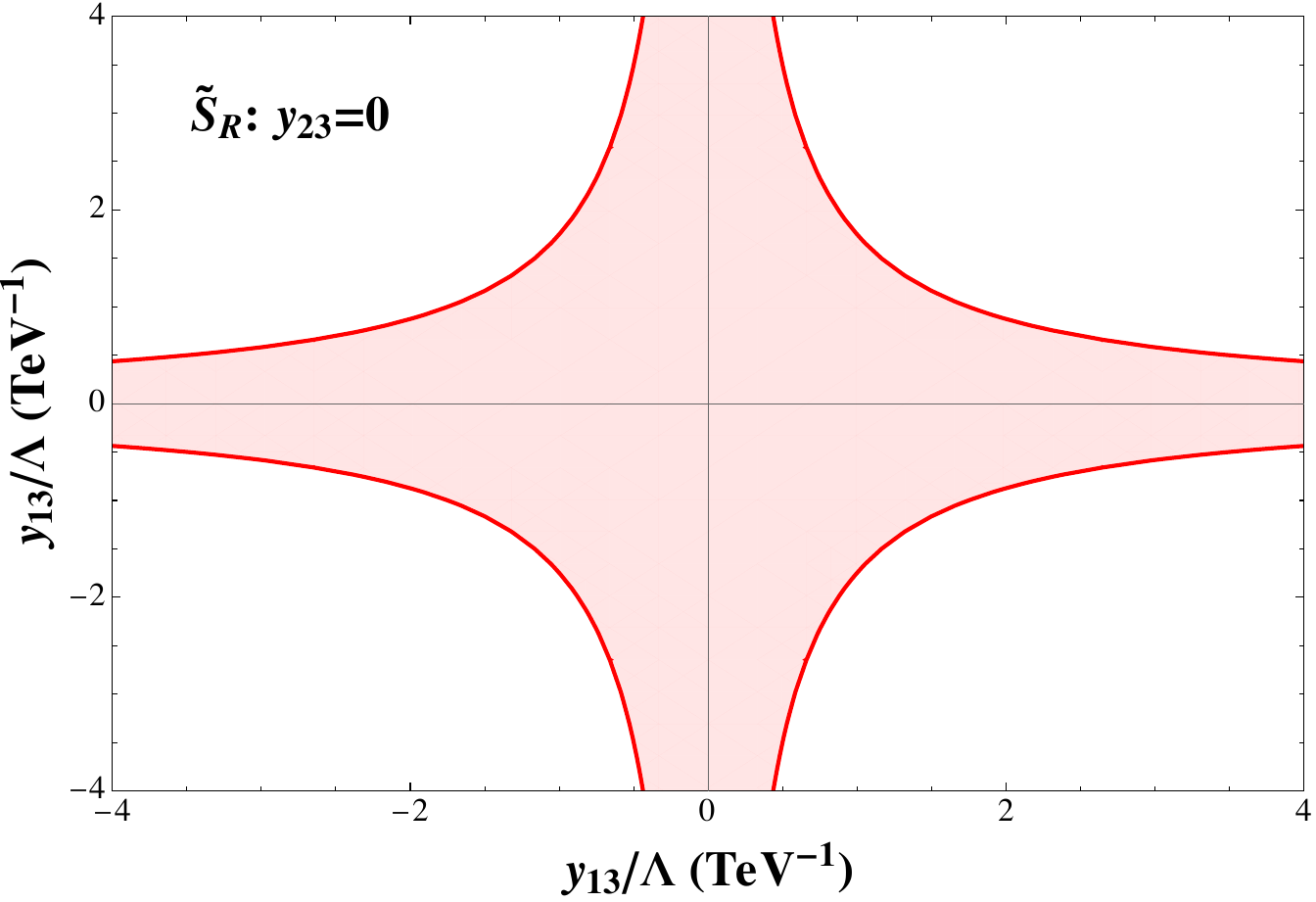}
\includegraphics[scale=0.25]{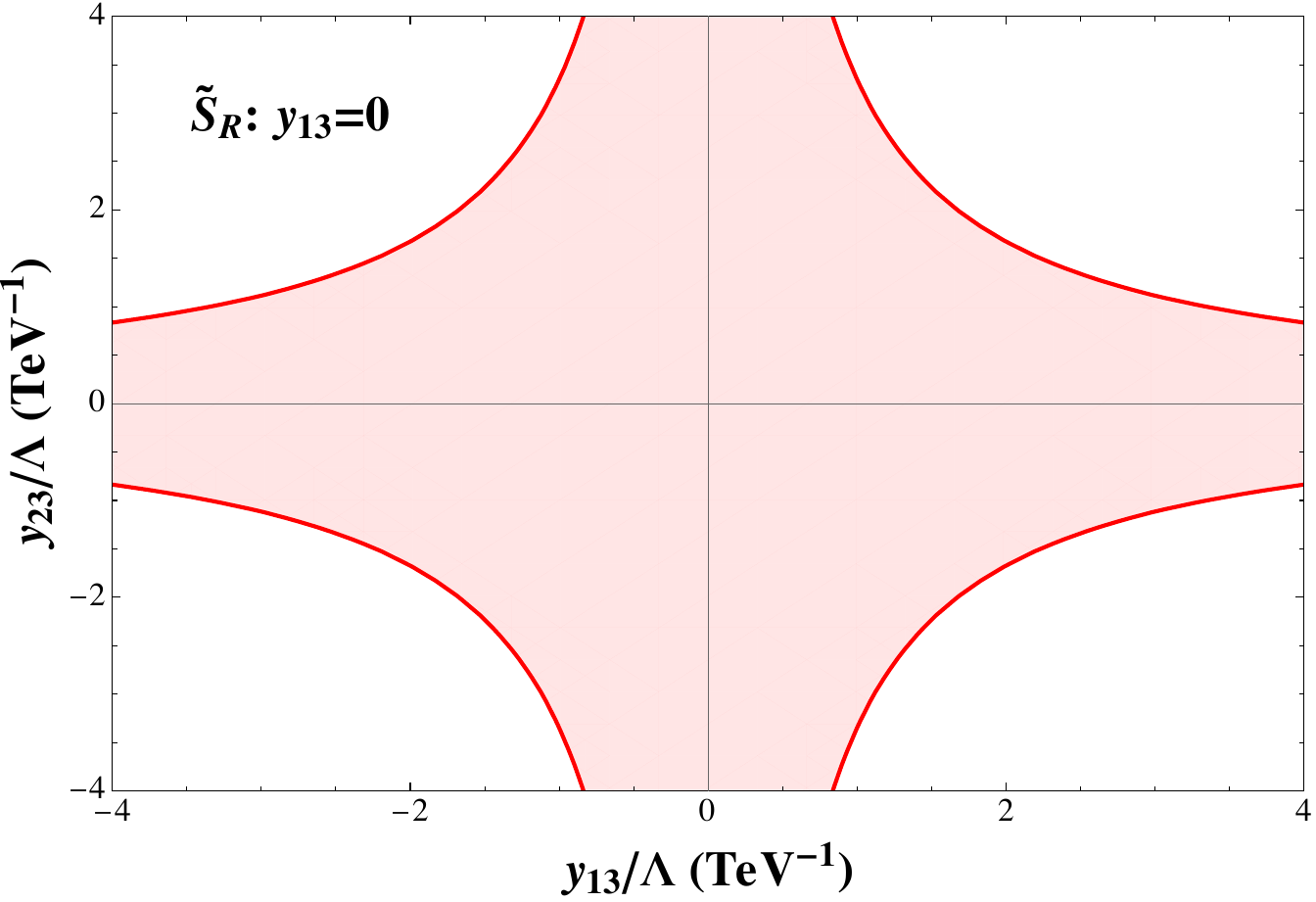}
\includegraphics[scale=0.25]{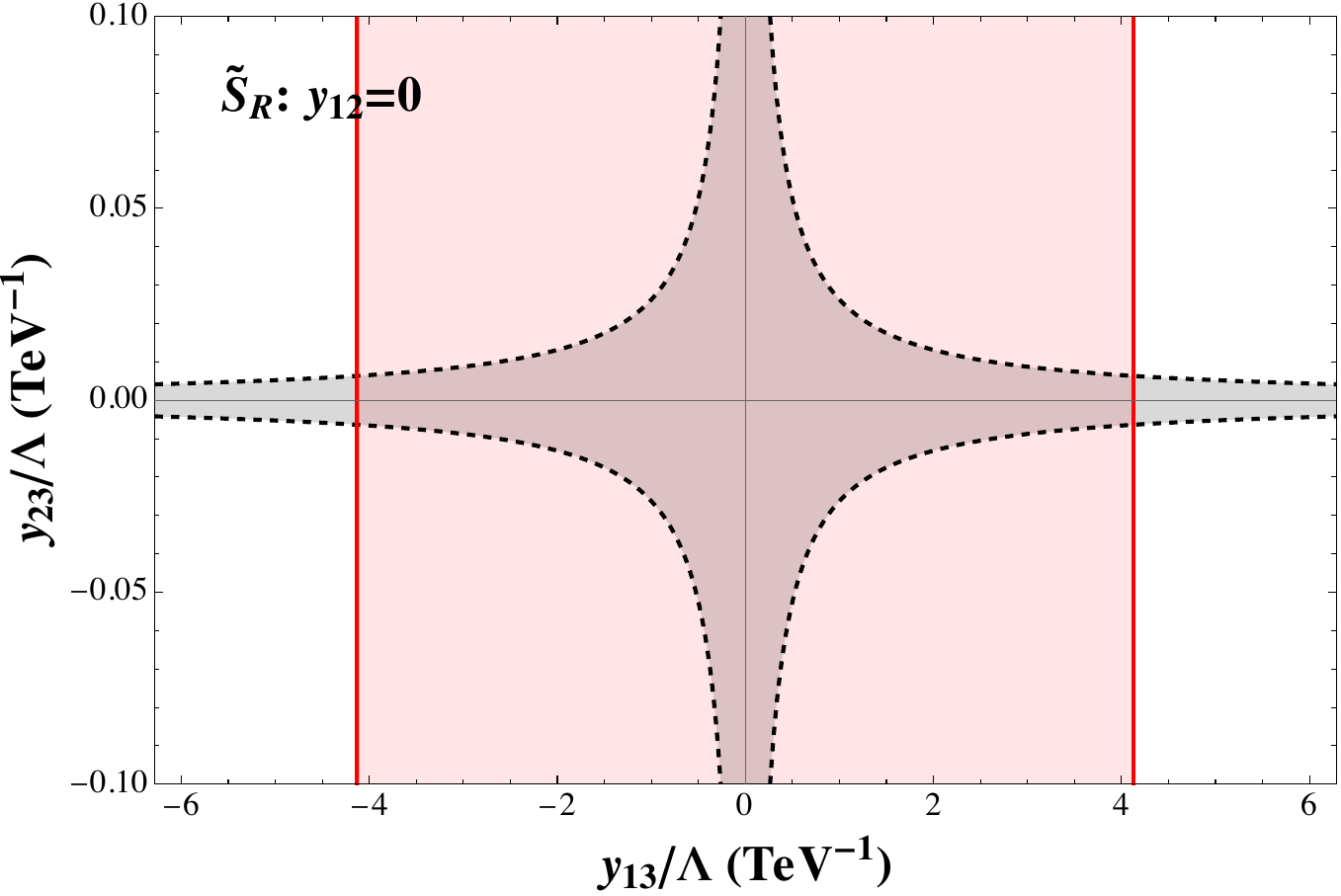}
\caption{The allowed regions in the parameter space for $\tilde S_R$. The light red region is given by the single top quark production cross section at 13TeV LHC. The gray region is given by the $D^0-\bar D^0$ mixing observable. The constraint from $D^0-\bar D^0$ mixing is not shown if it is always weak than the constraint from top physics. } 
\label{fig:sr}
\end{center}
\end{figure*}

The limit of the branching ratio of the top-quark rare decay processes are \cite{ATLAS:2023qzr,CMS:2021hug,ATLAS:2024mih}
\bea
{\text{Br}}(t\to Zq)&<&6.6\times 10^{-5},\\
{\text{Br}}(t\to hu)&<&1.9\times 10^{-4},\\
{\text{Br}}(t\to hc)&<&3.4\times 10^{-4}.
\eea
The numerical results show that these observables give weaker constraint in any case. Combining the constraints from the LHC and the $D^0-\bar D^0$ mixing, the allowed parameter region is shown in FIG. \ref{fig:zr}-\ref{fig:sr}.


\section{Conclusion}

In this work, we explored the landscape of heavy resonances $Z^\prime_R$, $G^\prime_R$ and $\tilde S_R$ which connect the right-handed top quark and the right-handed light up-quark directly in the UV model. The effective operators induced by these new particles at EW scale are generated. A preliminary analysis of the phenomenology is given. In FIG. \ref{fig:zr}-\ref{fig:sr}, it is shown clearly that  the phenomenology of $\tilde S_R$ is quite different from $Z^\prime_R$ and $G^\prime_R$. This should not be surprised since the fermion number violated resonance $\tilde S_R$ is more exotic than the fermion number conserved $Z^\prime_R$ and $G^\prime_R$, and so does not predict the very clear $tt$ signal at the LHC. So a signal of exotic single-top event without same-sign-top excess would favor the sextet scalar. And a signal of exotic same-sign-top event without single-top excess would favor the singlet and octet vector bosons. 

On the other hand, it is difficult to distinguish $Z^\prime_R$ and $G^\prime_R$. Although the effective operators and the Wilson coefficients look different in these two cases, one should remember that the interactions in the UV model have the same Lorentz structure. So the difference is from the color structure and color factor. To dig this information, the ratio $\delta\sigma(tt)/\delta(tj)$ is necessary. Nevertheless, more hard works are needed to suppress the theoretical uncertainty.

Our work also shows that the phenomenology can help us distinguish the pattern of the flavor-changing coupling constants in the UV model effectively. For $Z^\prime_R$ and $G^\prime_R$, whenever $y_{12}\neq 0$, it will contribute to the $D^0-\bar D^0$ mixing significantly and get strong constraint. Such property makes $D^0-\bar D^0$ mixing, associated with the same-sign-top signal at the LHC, be a powerful tool to discover these patterns of $Z^\prime_R$ and $G^\prime_R$. On the other hand, if people find excess from both the single-top and the same-sign-top event in near future, it would favor the $y_{12}=0$ pattern. For the color sextet, in the $y_{12}=0$ pattern, although the constraint from the $D^0-\bar D^0$ mixing is much weaker due to the loop suppression, it still gives valuable information since the other constraints are even weaker.

To distinguish the $y_{13}=0$ pattern and the $y_{23}=0$ pattern, one just needs to notice that the initial state partons are quite different in this two patterns. Such a difference does not only change the prediction of the total cross section, but also change the event shape since the existence of the valence $u$-quark tends to produce more energetic same-sign-top events via the $uu$ initial process, and a broader rapidity distribution of the single-top process via the $u\bar u$, $u\bar c$ and $u c$ initial processes.  

\begin{acknowledgments}
We would like to thank Dr. K. Chai and Dr. R. Zhang from the Theoretical Physics Division of the Institute of High Energy Physics (IHEP), Chinese Academy of Sciences, for their valuable assistance. The work of M. H and H. Z is supported in part by the National Science Foundation of China under Grants No. 12575114, No. 12235001 and No. 12075257. 
\end{acknowledgments}

\bibliography{main}
\end{document}